\documentclass[11pt]{article}   
\usepackage{amsmath,graphicx}
\usepackage{amssymb, amsxtra} 
\usepackage{blkarray}
\usepackage{mathtools}

\usepackage{graphicx}
\def\therefore{
\leavevmode
\lower0.1ex\hbox{$\bullet$}
\kern-0.2em\raise0.9ex\hbox{$\bullet$}
\kern-0.2em\lower0.2ex\hbox{$\bullet$}
\thinspace}

\title{A democratic suggestion}
\author{A. Kleppe, SACT, Oslo\\kleppe@nbi.dk}
\date{}
\begin{document}
\maketitle                                       

\begin{abstract}
Within the framework of quark mass matrices with a democratic texture, the unitary rotation matrices that diagonalize the quark matrices are obtained by a specific parametrization of the Cabibbo-Kobayashi-Maskawa mixing matrix. Different forms of democratic quark mass matrices are derived from slightly different parametrizations.
\end{abstract}

\section{Introduction}
The Standard Model is flawed by the large number of free parameters, for which there is at present no explanation.
There is no prediction of the family replication pattern, nor of the number of families. All the families are really treated on the same footing.
Most of the Standard Model free parameters reside in ``flavour space'' - with six quark masses, six lepton masses, four quark mixing angles and ditto for the leptonic sector, as well as the strong CP-violating parameter $\bar{\Theta}$. 
The structure of flavour space is determined by the fermion mass matrices, i.e. by the form that the mass matrices take in the ``weak interaction basis'' where mixed fermion states interact weakly, in contrast to the ``mass bases'', where the mass matrices are diagonal.

One may wonder how one may ascribe such importance to the different bases in flavour space, considering that
the information content of a matrix is contained in its matrix invariants, which in the case of a $N\times N$ matrix $M$ are the $N$ sums and products of the eigenvalues $\lambda_j$, such as $trace M$, $detM$,
\begin{equation}
  \def\arraystretch{1.1}
  \begin{array}{r@{\;}l} 
I_1 = &\sum_j\lambda_j                     = \lambda_1+\lambda_2+\lambda_3...\\

I_2 = &\sum_{jk}\lambda_j\lambda_k          = \lambda_1\lambda_2+\lambda_1\lambda_3+\lambda_1\lambda_4+... \\

I_3 = &\sum_{jkl}\lambda_j\lambda_k\lambda_l = \lambda_1\lambda_2\lambda_3+\lambda_1\lambda_2\lambda_4+...\\
      & \vdots\\
I_N = &\lambda_1\lambda_2 \cdots \lambda_N 
\end{array}
\end{equation} 
These expressions are invariant under permutations of the eigenvalues, which in the context of mass matrices means that they are flavour symmetric, and obviously independent of any choice of flavour space basis.

Even if the information content of a matrix is contained in its invariants,
the form of a matrix may also carry information, albeit of another type. The idea - the hope - is that the form that the mass matrices have in the weak interaction basis can give some hint about the origin of the unruly masses. There is a certain circularity to this reasoning; to make a mass matrix ansatz is in fact to define what we take as the weak interaction basis in flavour space.
We denote the quark mass matrices of the up- and down-sectors in the weak interaction basis by $M$ and $M'$, respectively. 
We go from the weak interaction basis to the mass bases by rotating the matrices by the unitary matrices $U$ and $U'$,
\begin{equation}\label{mss}
M \rightarrow UMU^{\dagger} = D = diag(m_u,m_c,m_t)
\end{equation}
\[
M' \rightarrow U'M'U'^{\dagger} = D' = diag(m_d,m_s,m_b)\\
\]
  \begin{figure}[htb]
    \begin{center}
    \includegraphics[scale=0.87]{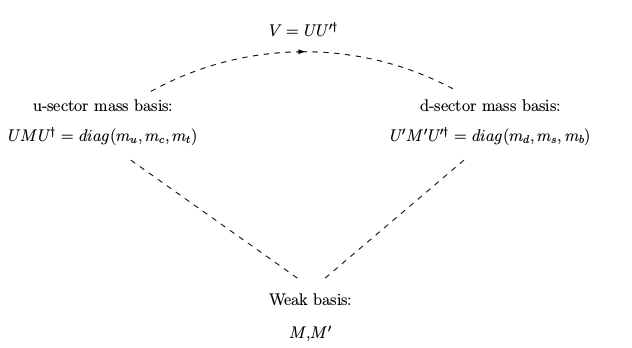}
\end{center}
    \end{figure}

The lodestar in the hunt for the right mass matrices is the family hierarchy, with two lighter particles in the first and second family, and a much heavier particle in the third family. This hierarchy is present in all the charged sectors, with fermions in different families exhibiting very different mass values, ranging from the electron mass to the about $10^{5}$ times larger top mass. It is still an open question whether the neutrino masses also follow this pattern \cite{neutrino hier}.

\section{Democratic mass matrices}
In the democratic approach  \cite{demo}, \cite{koide}, \cite{Fritzsch} the family hierarchy is taken very seriously. It is assumed that in the weak basis the fermion mass matrices have a form close to the $S(3)_L  \times  S(3)_R $ symmetric ``democratic'' matrix 
\begin{equation}\label{nambu}
{\bf{N}}= k\begin{pmatrix}
  1 & 1 & 1\\
  1 & 1 & 1\\
  1 & 1 & 1\\
\end{pmatrix}
\end{equation}
with the eigenvalues $(0,0,3k)$, reflecting the family hierarchy.

The underlying philosophy is that in the Standard Model, where the fermions get their masses from the Yukawa couplings by the Higgs mechanism, there is no reason why there should be a different Yukawa coupling for each fermion. 
The couplings to the gauge bosons of the strong, weak and
electromagnetic interactions are identical for all the fermions in a given charge sector, it thus seems like a natural assumption that they should also have identical Yukawa couplings.
The difference is that the weak interactions take place in a specific flavour space basis, while the other interactions are flavour independent. 

The democratic assumption is thus that the fermion fields of the same charge initially have the same Yukawa couplings. 
With three families, the quark mass matrices in the weak interaction basis then have the (zeroth order) form
\begin{equation}\label{nambu2}
M^{(0)}= k_u\begin{pmatrix}
  1 & 1 & 1\\
  1 & 1 & 1\\
  1 & 1 & 1\\
 \end{pmatrix},\hspace{2cm} 
M'^{(0)}= k_d\begin{pmatrix}
  1 & 1 & 1\\
  1 & 1 & 1\\
  1 & 1 & 1\\
 \end{pmatrix}
\end{equation}
where $k_u$ and $k_d$ have dimension mass. 
The corresponding mass spectra $(m_1,m_2,m_3) \sim (0,0,3k_j)$ reflect
the family hierarchy with two light families and a third much heavier family, a mass hierarchy that can be interpreted as the representation ${\bf{1}}\oplus {\bf{2}}$ of $S(3)$.
In order to obtain realistic mass spectra with non-zero masses, the $S(3)_L  \times  S(3)_R $ symmetry must obviously be broken, 
and the different democratic matrix ans\"{a}tze correspond to different schemes for breaking the democratic symmetry.

\subsection{The lepton sector}
We can apply the democratic approach to the lepton sector as well, postulating democratic (zeroth order) mass matrices for the charged leptons and the neutrinos, whether they are Fermi-Dirac or Majorana states, 
\begin{equation}\label{leptons}
M^{(0)}_l= k_l\begin{pmatrix}
  1 & 1 & 1\\
  1 & 1 & 1\\
  1 & 1 & 1\\
 \end{pmatrix},\hspace{2cm} 
M^{(0)}_{\nu}= k_{\nu}\begin{pmatrix}
  1 & 1 & 1\\
  1 & 1 & 1\\
  1 & 1 & 1\\
 \end{pmatrix}
\end{equation}
Relative to the quark ratio 
$k_u/k_d \sim m_t/m_b \sim  40 - 60$, the leptonic ratio $k_{\nu}/k_l < 10^{-8}$ is so extremely small that it seems unnatural. One way out is to simply assume that $k_{\nu}$ vanishes, meaning that the neutrinos get no mass contribution in the democratic limit \cite{Frit}. According to the democratic philosophy, then there would be no reason for a hierarchical pattern \`{a} la the one observed in the charged sectors; the neutrino masses could even be of the same order of magnitude.

Data are indeed compatible with a much weaker hierarchical structure for the neutrino masses than the hierarchy displayed by the charged fermion masses.

Unlike the situation for the quark mixing angles, in lepton flavour mixing there are two quite large mixing angles and a third much smaller mixing angle, these large mixing angles can be interpreted as indicating weak hierachy of the neutrino mass spectrum. The neutrino mass spectrum hierarchy could even be inverted; 
if the solar
neutrino doublet $(\nu_1,\nu_2)$ has a mean mass larger than the remaining
atmospheric neutrino $\nu_3$, the hierarchy is called
“inverted”, otherwise it is called “normal”. 

Supposing that the neutrino masses do not emerge from a democratic scheme, a (relatively) flat neutrino mass spectrum could be taken as a support for the idea that the masses in the charged sectors emerge from a democratic scheme.

\section{The democratic basis}
In the case that both the up- and down-sector mass matrices have a purely democratic texture, the quark mixing matrix is $V = UU'^{\dagger} = U_{dem}U_{dem}^{\dagger} = {\bf{1}}$, where  
\begin{equation}\label{dem}
U_{dem} =\frac{1}{\sqrt{6}}
 \begin{pmatrix}
  \sqrt{3} & -\sqrt{3} & \hspace{2mm} 0 \\
  1        &         1 & -2 \\
  \sqrt{2} & \sqrt{2}  & \sqrt{2}
 \end{pmatrix}
\end{equation}
is the unitary matrix that diagonalizes the democratic matrix (\ref{nambu}).
 
We use this to define the democratic basis, meaning the flavour space basis where the mass matrices are diagonalized by (\ref{dem}) and the mass Lagrangian is symmetric under permutations of the fermion fields $(\varphi_1,\varphi_2,\varphi_3)$ of a given charge sector.

In the democratic basis the mass Lagrangian
\[
{\mathcal{L}}_m =\bar{\varphi} M_{(dem)} \varphi =k\sum_{jk=1}^3 \bar{\varphi}_j \varphi_k
\]
is symmetric under permutations of the fermion fields $(\varphi_1,\varphi_2,\varphi_3)$, while 
in the mass basis
with 
\[
M_{(mass)} 
=\begin{pmatrix}
 \lambda_1\\
& \lambda_2\\
&& \lambda_3\\
\end{pmatrix}
\]
 the mass Lagrangian has the form
 \begin{equation} 
{\mathcal{L}}_m =\lambda_1 \bar{\psi}_1 \psi_1+\lambda_2 \bar{\psi}_2 \psi_2+\lambda_3 \bar{\psi}_3 \psi_3
\end{equation} 
which is clearly not invariant under permutations of ($\psi_1$,$\psi_3$,$\psi_3$). 

We can perform a shift of the democratic matrix, by just adding a unit matrix $diag(a,a,a)$, 
$M_0 \rightarrow M_1$, 
\begin{equation}\label{pert}
M_1 =
k \begin{pmatrix}
  1 & 1 & 1\\
  1 & 1 & 1\\
  1 & 1 & 1\\
 \end{pmatrix}+
\begin{pmatrix}
 a\\
&a\\
&&a\\
\end{pmatrix}=
\begin{pmatrix}
  k+a & k   & k\\
  k   & k+a & k\\
  k   & k   & k+a\\
 \end{pmatrix}
\end{equation}
corresponding to the mass spectrum $(a,a,3a+3k)$. The matrix
$M_1$ has a democratic texture, both because it is diagonalized by $U_{dem}$, and because the mass Lagrangian is invariant under permutations of the quark fields,
\begin{equation}{\mathcal{L}}_{M_1} = (k+a)\sum \bar{\varphi}_j \varphi_j+ k \sum_{j\neq k} \bar{\varphi}_j \varphi_j
\end{equation}

If $M_1$ and $M'_1$ both have a texture like (\ref{pert}), there is no CP-violation. This is independent of how many families there are, because of the degeneracy of the mass values. 
CP-violation only occurs once there are three or more non-degenerate families, because only then the phases can no longer be defined away.

We can repeat the democratic scheme with a number $n$ of families, where the fermion mass matrices again are proportional to the $S(n)_L  \times  S(n)_R $ symmetric democratic matrix 
which is diagonalized by a unitary matrix analogous to $U_{dem}$ in (\ref{dem}). 
To the $n\times n$-dimensional democratic matrix term, we can again add a $n\times n$-dimensional diagonal matrix $diag(a,a,...,a)$, and get a $n\times n$-dimensional mass spectrum 
with $n$ massive states, and $n-1$ degenerate masses.
The mass matrix still has a democratic texture, and there is still no CP-violation.

\section{Breaking the democratic symmetry}
In order to obtain non-degenerate, non-vanishing masses for the physical flavours $(\psi_1,\psi_2,\psi_3)$, the permutation symmetry of the fermion fields $(\varphi_1,\varphi_2,\varphi_3)$ in the democratic basis must be broken. The proposal here is to derive mass matrices with a nearly democratic texture, not by explicitly perturbing the assumed initial democratic form (\ref{nambu}), but instead by perturbing the matrix $U_{dem}$ which diagonalises the democratic mass matrix. This is done by deriving the unitary rotation matrices $U$, $U'$ for the up- and down- sectors, from a specific parameterisation of the weak mixing matrix $V = UU^{'\dagger}$.

The idea is to embed the assumption of democratic symmetry into the Standard Model mixing matrix, by expressing the mixing matrix as a product
\begin{equation}\label{UU}  
V=UU'^{\dagger}=  ({\tilde{U}}U_{dem})(U_{dem}^{\dagger}{\tilde{U'}}^{\dagger})
\end{equation}  
Since both the mixing matrix and its factors, according to the standard parameterisation \cite{Stand}, are so close to the unit matrix, the rotation matrices $U$, $U'$ are effectively perturbations of the democratic diagonalising matrix (\ref{dem}). In this way, the weak interaction basis remains close to the democratic basis.

\subsection{Factorizing the mixing matrix}   
The Cabbibo-Kobayashi-Maskawa (CKM) mixing matrix \cite{CKM} can of course be parametrized - and factorized - in many different ways, and different factorizations correspond to different
rotation matrices $U$ and $U'$.
The most obvious and ``symmetric'' factorization of the CKM mixing matrix is, following the standard parametrization \cite{Stand} with three Euler angles $\alpha$, $\beta$, $2\theta$,
\begin{equation}
V =\begin{pmatrix}
  c_{\beta} c_{2\theta}                                     &s_{\beta} c_{2\theta}                                     &s_{2\theta} e^{-i\delta}\\
-c_{\beta} s_{\alpha} s_{2\theta} e^{i\delta}-s_{\beta}c_{\alpha}&-s_{\beta}s_{\alpha}s_{2\theta} e^{i\delta}+c_{\beta}c_{\alpha}&s_{\alpha}c_{2\theta}\\
-c_{\beta} c_{\alpha} s_{2\theta} e^{i\delta}+s_{\beta}s_{\alpha}&-s_{\beta}c_{\alpha}s_{2\theta} e^{i\delta}-c_{\beta}s_{\alpha}&c_{\alpha}c_{2\theta}\\
\end{pmatrix}= UU^{'\dagger}
\end{equation} 
with the diagonalizing rotation matrices for the up- and down-sectors 
\begin{equation}\label{diagu}
U =
 \begin{pmatrix}
  1 & 0          & 0 \\
  0 &\cos\alpha  &\sin\alpha \\
  0 &-\sin\alpha &\cos\alpha \\
 \end{pmatrix}
\begin{pmatrix}
  e^{-i\gamma} \\
  & 1\\
  && e^{i\gamma} \\
 \end{pmatrix}
\begin{pmatrix}
  \cos\theta     & 0          & \sin\theta \\
  0              & 1          & 0\\
  -\sin\theta    & 0          & \cos\theta \\
 \end{pmatrix}
 \begin{pmatrix}
 \frac{1}{\sqrt{2}}  & -\frac{1}{\sqrt{2}} & \hspace{2mm} 0 \\
  \frac{1}{\sqrt{6}} &\frac{1}{\sqrt{6}}   & -\frac{2}{\sqrt{6}} \\
  \frac{1}{\sqrt{3}} &\frac{1}{\sqrt{3}}   & \frac{1}{\sqrt{3}}
 \end{pmatrix}
\end{equation}
and
\[
U' =
 \begin{pmatrix}
  \cos\beta  &-\sin\beta &0\\
  \sin\beta  &\cos\beta  &0\\
   0         & 0         &1
 \end{pmatrix}
\begin{pmatrix}
  e^{-i\gamma} \\
  & 1\\
  && e^{i\gamma} \\
 \end{pmatrix}
\begin{pmatrix}
  \cos\theta     & 0          & -\sin\theta \\
  0              & 1          & 0\\
   \sin\theta    & 0          & \cos\theta \\
 \end{pmatrix}
 \begin{pmatrix}
 \frac{1}{\sqrt{2}}  & -\frac{1}{\sqrt{2}} & \hspace{2mm} 0 \\
  \frac{1}{\sqrt{6}} &\frac{1}{\sqrt{6}}   & -\frac{2}{\sqrt{6}} \\
  \frac{1}{\sqrt{3}} &\frac{1}{\sqrt{3}}   & \frac{1}{\sqrt{3}}
 \end{pmatrix},
\]
respectively,
where $\alpha$, $\beta$, $\theta$ and $\gamma$ correspond to the parameters in the standard parametrization in such a way that
$\gamma = \delta/2$, $\delta = 1.2 \pm 0.08$ rad, and $2\theta = 0.201 \pm 0.011 ^{\circ}$, while  
$\alpha = 2.38 \pm 0.06 ^{\circ}$ 
and $\beta = 13.04\pm 0.05 ^{\circ}$.

From the rotation matrices $U$ and $U'$ we then obtain the mass matrices 
$M={U^{\dagger}}diag(m_u,m_c,m_t)U$ and 
$M'={U'^{\dagger}}diag(m_d,m_s,m_b)U'$, such that  
\begin{equation}\label{1}
M=\frac{1}{6}\begin{pmatrix}
X+H                        &\hat{M}_{12}                    &Z+W\\
\hat{M}_{12}^*\hspace{0.5mm}      &X-H                      &Z-W\\
\hspace{0.5mm}Z^*+W^*      &\hspace{2mm}Z^*-W^*      &6T-2X\\
  \end{pmatrix} 
\end{equation}
where $T$ is the trace $T= m_u+m_c+m_t$, and with $D= \sqrt{3}s_{\theta}  -\sqrt{2}c_{\theta}$, 
$ C= \sqrt{3}s_{\theta}  +\sqrt{2}c_{\theta}$, 
$ F= c_{\alpha}s_{\alpha}(m_t-m_c)$,
\begin{description}
\item $ X=\frac{1}{2}(m_cs_{\alpha}^2+m_tc_{\alpha}^2-m_u)(D^2+C^2-2)+F(D-C)\cos\gamma+T+3m_u$
\item $ H=\frac{1}{2}(m_cs_{\alpha}^2+m_tc_{\alpha}^2-m_u)(D^2-C^2)+F\cos\gamma(D+C)$
\end{description}
\begin{description}
\item $W = \frac{1}{4}(m_cs_{\alpha}^2+m_tc_{\alpha}^2-m_u)\hspace{1mm}(D^2-C^2)-F\hspace{1mm}(D+C)\hspace{1mm}e^{-i\gamma}$
\item $Z = (m_cs_{\alpha}^2+m_tc_{\alpha}^2-m_u)\left[2+\frac{1}{4}(D-C)^2\right]\hspace{1mm}+\frac{F}{2}\hspace{1mm}(D-C)\hspace{1mm}(e^{i\gamma}-2\hspace{1mm}e^{-i\gamma})-2T+6\hspace{1mm}m_u$
\item $\hat{M}_{12}= -(m_cs_{\alpha}^2+m_tc_{\alpha}^2-m_u)\hspace{1mm}(D\hspace{1mm}C+1)-F\hspace{1mm}(C\hspace{1mm}e^{i\gamma}-D\hspace{1mm}e^{-i\gamma})+T-3\hspace{1mm}m_u$ 
\end{description}
Similarly for the down-sector,
\begin{equation}\label{11}
M'=\frac{1}{6}\begin{pmatrix}
X'+H'                        &\hat{M}'_{12}                    &Z'+W'\\
\hat{M}_{12}^{'*}\hspace{0.5mm}      &X'-H'                      &Z'-W'\\
\hspace{0.5mm}Z^{'*}+W^{'*}      &\hspace{2mm}Z^{'*}-W^{'*}      &6T'-2X'\\
  \end{pmatrix} 
\end{equation}
with the parameters $T'= m_d+m_s+m_b$, $ G= \sqrt{2}s_{\theta}  -\sqrt{3}c_{\theta}$, $ J= \sqrt{2}s_{\theta}  +\sqrt{3}c_{\theta} $,
\begin{description}
\item$ F'= c_{\beta}s_{\beta}(m_b-m_s)$, and 
\item $X'= \frac{1}{2}(m_ss_{\beta}^2+m_bc_{\beta}^2-m_d)(G^2+J^2-2)-F'(J+G)\cos\gamma+T'+3m_b$
\item $H'= \frac{1}{2}(m_ss_{\beta}^2+m_bc_{\beta}^2-m_d)(G^2-J^2)+F'(J-G)\cos\gamma$
\item $W'= \frac{1}{4}(m_ss_{\beta}^2+m_bc_{\beta}^2-m_d)(G^2-J^2)+F'(G-J)e^{i\gamma}$
\item $Z'=(m_ss_{\beta}^2+m_bc_{\beta}^2-m_d)\left[2+\frac{1}{4}(J+G)^2\right]+\frac{F'}{2}(J+G)(2e^{i\gamma}-e^{-i\gamma})-2T'+6m_b$
\item $\hat{M}'_{12}= (m_ss_{\beta}^2+m_bc_{\beta}^2-m_d)\hspace{1mm}(G\hspace{1mm}J-1)-F'\hspace{1mm}(J\hspace{1mm}e^{i\gamma}-G\hspace{1mm}e^{-i\gamma})+T'-3\hspace{1mm}m_b$ 
\end{description}

In order to evaluate to what degree these rather opaque matrices are democratic, we calculate numerical matrix elements by inserting numerical mass values.
For the up-sector
we get the (nearly democratic) matrix texture
\begin{equation}\label{CU}
M=C_u\left[\begin{pmatrix}
  1\\ 
  & k\hspace{1mm}e^{-i(\mu+\rho)} \\
  &&kp\hspace{1mm} e^{-i\mu} \\
 \end{pmatrix}
\begin{pmatrix}
1&1&1\\
1&1&1\\
1&1&1\\
\end{pmatrix}
\begin{pmatrix}
  1\\
  & k\hspace{1mm}e^{i(\mu+\rho)} \\
  &&kp\hspace{1mm} e^{i\mu} \\
\end{pmatrix}+\Lambda \right]
\end{equation}
where the ``small'' matrix
\[
\Lambda=
\begin{pmatrix}
0   &                   0&   0\\
0   &\varepsilon         &\varepsilon'e^{-i\rho}\\
0   &\varepsilon'e^{i\rho}&\eta\\
\end{pmatrix},
\]
with 
$\varepsilon \sim \varepsilon' \ll \eta < k, p$, is what breaks the democratic symmetry, supplying the two lighter families with non-zero masses. With mass values calculated at $\mu = M_Z$ (Jamin 2014) \cite{Jamin}, 
\begin{equation}\label{jam1}
(m_u(M_Z),m_c(M_Z),m_t(M_Z)) =(1.24, 624 , 171550 ) MeV,
\end{equation}
we get 
\begin{description}
\item$\mu \sim 2.7895^o$,\hspace{1mm}  $\rho \sim 2.7852^o$,\hspace{1mm}$C_u=54240.36$ MeV $\approx m_t/3 $,
and
\item$k\approx 1.00438$,\hspace{1mm} $p\approx 1.06646$,\hspace{1mm}$\varepsilon' \approx 5.05\hspace{1mm} 10^{-5}$,
\item$\varepsilon \approx 4.6\hspace{1mm} 10^{-5} \approx 2\frac{m_u}{C_u}$,\hspace{1mm}$\eta = 1.815\hspace{1mm} 10^{-2}\approx \frac{1}{2}\frac{m_t}{C_u}\frac{m_c}{C_u}$.
\end{description}
For the down-sector, with 
\begin{equation}\label{jam2}
(m_d(M_Z), m_s(M_Z), m_b(M_Z))=(2.69, 53.8, 2850) MeV
\end{equation}
we get another democratic texture,
\begin{equation}
M'=C_d
\begin{pmatrix}
X+A  &  Ye^{-i\tau}   &\hspace{2mm} e^{-i\nu}\\    
Ye^{i\tau}   &  X-A  & (1+2A)e^{i\kappa}\\    
e^{i\nu}   & (1+2A)e^{-i\kappa}  &\hspace{2mm} X+Y-A-1\\    
\end{pmatrix}
\end{equation}
where 
\begin{description}
\item$C_d=966.5 MeV$, $A=5.6\hspace{1mm}10^{-3}$, $X=1.0362$, $Y=1.0305$, and
$\tau \leq \kappa \sim 0.22^o$ $<$ $\nu \sim 0.226^o$.
\end{description}
Just like in the up-sector mass matrix, the matrix elements in $M'$ display a nearly democratic texture. For both the up-sector and the down-sector the mass matrices are thus approximately democratic.

\section{Calculability}
In the mass matrix literature there is an emphasis on ``calculability''. The ideal is to obtain mass matrices that have a manageable form, but there is nothing that forces nature to serve us such user-friendly formalism. It is however tempting to speculate that there are relations between the elements that could make the democratic matrices more calculable, and in the search for matrices that are reasonably transparent and calculable, we look at a more radical factorization of the mixing matrix, viz. 
\begin{equation}\label{ddiagun}
U =
 \begin{pmatrix}
  1 & 0          & 0 \\
  0 &\cos\alpha  &\sin\alpha \\
  0 &-\sin\alpha &\cos\alpha \\
 \end{pmatrix}
\begin{pmatrix}
  \cos\omega                & 0          & \sin\omega \hspace{1mm}e^{-i\delta} \\
  0                         & 1          & 0\\
  -\sin\omega \hspace{1mm}e^{ i\delta}    & 0          & \cos\omega \\
 \end{pmatrix}
 \begin{pmatrix}
 \frac{1}{\sqrt{2}}  & -\frac{1}{\sqrt{2}} & \hspace{2mm} 0 \\
  \frac{1}{\sqrt{6}} &\frac{1}{\sqrt{6}}   & -\frac{2}{\sqrt{6}} \\
  \frac{1}{\sqrt{3}} &\frac{1}{\sqrt{3}}   & \frac{1}{\sqrt{3}}
 \end{pmatrix}
\end{equation}
and
\[
U' =
 \begin{pmatrix}
  \cos\beta  &-\sin\beta &0\\
  \sin\beta  &\cos\beta  &0\\
   0         & 0         &1
 \end{pmatrix}
 \begin{pmatrix}
 \frac{1}{\sqrt{2}}  & -\frac{1}{\sqrt{2}} & \hspace{2mm} 0 \\
  \frac{1}{\sqrt{6}} &\frac{1}{\sqrt{6}}   & -\frac{2}{\sqrt{6}} \\
  \frac{1}{\sqrt{3}} &\frac{1}{\sqrt{3}}   & \frac{1}{\sqrt{3}}
 \end{pmatrix}
\]
where, as before, $\delta = 1.2 \pm 0.08$ rad, and $\omega=2\theta = 0.201 \pm 0.011 ^{\circ}$, while  
$\alpha = 2.38 \pm 0.06 ^{\circ}$, 
and $\beta = 13.04\pm 0.05 ^{\circ}$.
These rotation matrices are still ``perturbations'' of the democratic diagonalizing matrix (\ref{dem}), and
the up-sector mass matrix has a texture similar to (\ref{1}),
\begin{equation}\label{4}
M=
\frac{1}{6}\begin{pmatrix}
R+Q+S\hspace{1mm}\cos\delta             &R-Q-iS\hspace{1mm}\sin\delta& A-Be^{-i\delta}\\
R-Q+iS\hspace{1mm}\sin\delta&R+Q-S\hspace{1mm}\cos\delta             & A+Be^{-i\delta}\\
A-Be^{i\delta}                &A+Be^{i\delta}               &T-2(R+Q)\\
  \end{pmatrix} 
\end{equation}
where $T$ is the trace, $T= m_u+m_c+m_t$, and
\begin{description}
\item $R= N \hspace{1mm}(2\hspace{1mm}c_{\omega}^2-1)+T-2\hspace{1mm}\sqrt{2}\hspace{1mm}c_{\omega}\hspace{1mm}F$, \hspace{1mm} $Q= 3\hspace{1mm}s_{\omega}^2\hspace{1mm}N+3\hspace{1mm}m_u$, 
\item$S= -2\sqrt{6}\hspace{1mm}c_{\omega}\hspace{1mm}s_{\omega}\hspace{1mm}N+2\hspace{1mm}\sqrt{3}s_{\omega}\hspace{1mm}F$ 
\item $A= N\hspace{1mm}(2\hspace{1mm}c_{\omega}^2+2)-2\hspace{1mm}T+\sqrt{2}\hspace{1mm}c_{\omega}\hspace{1mm}F+6\hspace{1mm}m_u$, \hspace{1mm} $B= \sqrt{6}\hspace{1mm}c_{\omega}\hspace{1mm}s_{\omega}\hspace{1mm}N+2\hspace{1mm}\sqrt{3}\hspace{1mm}F\hspace{1mm}s_{\omega}$ 
\end{description}
with
$N= m_c\hspace{1mm}s_{\alpha}^2+m_t\hspace{1mm}c_{\alpha}^2-m_u$,\hspace{1mm} $F= c_{\alpha}\hspace{1mm}s_{\alpha}\hspace{1mm}(m_t-m_c)$.
This matrix can be reformulated in a form similar to (\ref{CU}), 

\[ 
M_u=
C_u \left[\begin{pmatrix}
  1\\
  & k\hspace{1mm}e^{-i\mu} \\
  &&kp\hspace{1mm}e^{-i(\mu-\rho)}  \\
 \end{pmatrix}
\begin{pmatrix}
1&1&1\\
1&1&1\\
1&1&1\\
\end{pmatrix}
\begin{pmatrix}
  1\\
  & k\hspace{1mm}e^{i\mu} \\
  &&kp\hspace{1mm}e^{i(\mu-\rho)} \\ 
\end{pmatrix}+\Lambda\right]
\]
where $C_u=R+Q+S\cos\delta$, $\mu=\arctan\left[S\sin\delta/(Q-R)\right]$, $\rho=\arctan\left(B\sin\delta/(A+B\cos\delta)\right)$, and
\[
\Lambda=
\begin{pmatrix}
0   &                   0&   0\\
0   &\varepsilon         &\varepsilon'e^{-i\rho}\\
0   &\varepsilon'e^{i\rho}&\eta\\
\end{pmatrix}
\]
with 
\begin{description}
\item $k=|M_{12}|/M_{11}=\frac{|R-Q-iS\sin\delta|}{R+Q+S\cos\delta}$,\hspace{4mm}$p=|M_{13}|/|M_{12}|=\frac{|A-Be^{-i\delta}|}{|R-Q-iS\sin\delta|}$,
\item $\varepsilon=(|M_{22}||M_{11}|-|M_{12}|^2)/|M_{11}|^2=
\frac{4RQ-S^2}{|R+Q+S\cos\delta|^2}$,
\item $\varepsilon'=(|M_{23}||M_{11}|-|M_{13}||M_{12}|)/|M_{11}|^2$,
\item$\eta=(|M_{33}||M_{11}|-|M_{13}|^2)/|M_{11}|^2$
\end{description}
Inserting the mass values (\ref{jam1}) gives
\begin{description}
\item$C_u=53723.5 MeV$,\hspace{1mm}$k=1.00318$,\hspace{1mm} $p=1.0828$, and
\item $\varepsilon\approx4.65\hspace{1mm}10^{-5} \approx 2\frac{m_u}{C_u}$,\hspace{1mm} $\varepsilon'\approx4.44\hspace{1mm}10^{-5}$,\hspace{1mm} $\eta\approx1.85\hspace{1mm} 10^{-2}\approx \frac{1}{2}\frac{m_t}{C_u}\frac{m_c}{C_u}$
\end{description}
For the down-sector, with 
\[  
U' =
 \begin{pmatrix}
  \cos\beta  &-\sin\beta &0\\
  \sin\beta  &\cos\beta  &0\\
   0         & 0         &1
 \end{pmatrix}
 \begin{pmatrix}
 \frac{1}{\sqrt{2}}  & -\frac{1}{\sqrt{2}} & \hspace{1mm} 0 \\
  \frac{1}{\sqrt{6}} &\frac{1}{\sqrt{6}}   & -\frac{2}{\sqrt{6}} \\
  \frac{1}{\sqrt{3}} &\frac{1}{\sqrt{3}}   & \frac{1}{\sqrt{3}}
 \end{pmatrix},
\]
the mass matrix $U'^{\dagger} diag(m_d,m_s,m_b)U'$ reads 
\[
M'=
C_d\begin{pmatrix}
X+A  &  Y   &\hspace{1mm} 1\\    
Y   &  X-A  &\hspace{1mm} 1+2A\\    
1   & 1+2A  &\hspace{1mm} X+Y-A-1\\    
\end{pmatrix}
\]
where
\begin{description} 
\item$C_d= 2(m_d c^2_{\beta}+m_s s^2_{\beta})-2\sqrt{3}c_{\beta}s_{\beta}(m_s-m_d))+2(m_b-m_s-m_d)$
\item$X= (2m_b+m_s+m_d+2(m_d c^2_{\beta}+m_s s^2_{\beta})+2\sqrt{3}c_{\beta}s_{\beta}(m_s-m_d))/C_d$
\item$Y=(2m_b+m_s+m_d-4(m_d c^2_{\beta}+m_s s^2_{\beta}))/C_d$,
\item$A=2\sqrt{3}c_{\beta}s_{\beta}(m_s-m_d))/C_d$. 
\end{description}
Inserting the mass values (\ref{jam2}) 
we moreover get the numerical values 
\begin{description}
\item $C_d= 926.448 MeV \approx m_b/3$,\hspace{1mm}$X = 1.0375$,\hspace{1mm}$A = 7\hspace{1mm} 10^{-3}$,\hspace{1mm}$Y = 1.0318$.
\end{description}

\section{Conclusion}
By including the democratic rotation matrix in the parametrization of the weak mixing matrix, we obtain mass matrices with specific democratic textures. In this way we make contact between the democratic hypothesis and the experimentally derived parameters of the CKM mixing matrix, avoiding the introduction of additional concepts.

\end{document}